\documentclass[conference]{IEEEtran}
\usepackage[T1]{fontenc}
\IEEEoverridecommandlockouts
\usepackage{amsmath,amssymb,amsfonts}
\usepackage{algorithmic}
\usepackage{graphicx}
\usepackage{subcaption}
\usepackage{textcomp}
\usepackage{xcolor}
\usepackage{braket}
\usepackage{dsrm}
\usepackage{url}
\usepackage{booktabs}
\usepackage{multirow}
\usepackage{tikz}
\usetikzlibrary{quantikz2}
\usetikzlibrary{shapes.multipart, arrows.meta}
\usepackage[frozencache,cachedir=.]{minted}

\begin{document}

\title{

Systematic Experiment Tracking in Quantum Software: A Case Study of Reservoir Computing with Error Mitigation
\thanks{This work has been supported by Business Finland (EM4QS 155/31/2024), Finnish Ministry of Education and Culture through the Quantum Doctoral Education Pilot Program (QDOC VN /3137/2024-OKM-4) and the Research Council of Finland through Finnish Quantum Flagship project (359240, JYU).}
}

\author{\IEEEauthorblockN{Otso Kinanen}
\IEEEauthorblockA{\textit{University of Jyväskylä}\\
Jyväskylä, Finland}
\and
\IEEEauthorblockN{Valter Uotila}
\IEEEauthorblockA{\textit{Aalto University}\\
Espoo, Finland}
\and
\IEEEauthorblockN{Vlad Stirbu}
\IEEEauthorblockA{\textit{University of Jyväskylä}\\
Jyväskylä, Finland}
}

\maketitle

\begin{abstract}
Quantum computers are more widely available than ever, making the field more accessible and widespread. Practitioners are coming from a wide range of domains, conducting experiments and research using quantum computing approaches across a variety of problems. The current literature suggests that developers follow certain methodologies in quantum software development, often with a matching set of tools provided. Yet with the novel paradigm, there are areas that remain unaddressed in practices and tools. In this article, we go into the details of experiment tracking in quantum software development. We explain the basic concept of experiment tracking and detail how, in essence, quantum computing sets demands on tracking practices. Given the experimental state of hardware and the constantly evolving software, quantum execution must be monitored, marginal gains aggregated for the best outcome, and error sources detected. In our case study, quantum reservoir computing for chaotic time series data prediction with error mitigation, we present a detailed quantum software development process and describe how experiments can be tracked throughout development. We then generalize this knowledge into the broader quantum development process. 
\end{abstract}


\section{Introduction}

As quantum computing is evolving from proof-of-concept demonstrations towards practical applications, systematic experiment tracking will become a critical enabler for reproducible experiments in academia and industry. With new hardware milestones reached~\cite{Abanin_Acharya_Aghababaie-Beni_Aigeldinger_Ajoy_Alcaraz_Aleiner_Andersen_Ansmann_Arute, Acharya_Abanin_Aghababaie-Beni_Aleiner_Andersen_Ansmann_Arute_Arya_Asfaw_Astrakhantsev_et,Arute_Arya_Babbush_Bacon_Bardin_Barends_Biswas_Boixo_Brandao_Buell_et,Kim_Eddins_Anand_Wei_vandenBerg_Rosenblatt_Nayfeh_Wu_Zaletel_Temme,doi:10.1126/sciadv.adu9991} and advancements in quantum software~\cite{unitary2025qoss,10646543}, many experts believe we are steadily on the way to reaching various quantum advantages~\cite{Eisert_Preskill_2025,Lanes_Beji_Corcoles_Dalyac_Gambetta_Henriet_Javadi-Abhari_Kandala_Mezzacapo_Porter_et}. With the growing number of practitioners in quantum computing, the demand for suitable tooling has increased in parallel. Especially as the development scope moves towards larger scales, the demand for methods and supporting tools increases with code complexity. Larger qubit counts and lower error rates in recent devices enable researchers to broaden their experimentation, execute longer circuits, incorporate more complex data into experiments, and explore possibilities in areas such as error mitigation and even error correction~\cite{Abanin_Acharya_Aghababaie-Beni_Aigeldinger_Ajoy_Alcaraz_Aleiner_Andersen_Ansmann_Arute}. Unfortunately, this does not yet mean that errors and other hardware limitations have been eliminated, or that they should not be addressed during development. The challenges require iterative quantum software development cycles, which means that experiments must be tracked systematically and automatically~\cite{kinanen2025toolchain}.

The quantum software development process and toolkits should take these challenges into consideration and support users in achieving the best possible outcomes from the development and execution processes~\cite {murillo2025quantum}. However, existing experiment tracking frameworks are designed primarily for classical machine learning workflows and lack the specialized features that are unique to quantum software development, such as circuit-related parameters, measurement result handling, and noise characteristics. Comprehensive tracking of experiments is the key to enabling robust reproducibility, and this demands that the experiment tracking be extended and improved from its classical state. In quantum software development, tracking experiments involves following methods verified for the purpose and selecting tools accordingly. In quantum execution, this means considering the exact platform in use and tracking execution metrics for each run. Doing this will allow the developer to better analyze and verify results obtained from their experiments, but it may not always be trivial, and practices might depend on the hardware platform, software development kit, or many other factors.

In our article, we have used the design science research methodology with an objective-centered approach~\cite{peffers2007design}. Our central research objective is the following:

\begin{quote}
\textit{How can we assist quantum practitioners in collecting experimental data and improving decision-making for better software engineering?}
\end{quote}

With the design science research methodology, we investigate the development process, specifically the experiment tracking as a supporting activity, covering both practices and tools. We explain how to improve the development process by tracing data across development phases and analyzing experimental results. We demonstrate the work with a case study – quantum reservoir computing with error mitigation – showing how, in this case, the experiment tracking supported and improved the development process, while providing better reproducibility in a scientific context. The experiment tracking tool used in our experiment is MLflow, which allows us to store the necessary data and evaluate the results and progress during development. We also identify experiment-tracking dimensions that are characteristic of quantum computing and have not previously been observed in classical machine learning pipelines. These dimensions require special features from experiment tracking tools and introduce demand for improving and adjusting the existing tools for quantum software development pipelines.

Considering the selected case study, we present a realistic quantum software development pipeline that employs quantum reservoir computing to predict chaotic time-series data. Quantum reservoir computing is a non-variational quantum machine learning paradigm that we consider especially suitable for the case study, as it allows us to demonstrate many of the realistic steps in quantum software development. These steps include classical data preparation, noisy execution, an advanced state-tomography-based measurement scheme, error mitigation with tensor networks, classical post-processing and readout, and finally cross-validation of the results between various approaches. These features create a compelling and realistic case study that requires tracking multiple parameters and metrics throughout the process.

We present the following contributions in this article:
\begin{itemize}
    \item We introduce and define experiment tracking practices in the quantum software development context.
    \item We integrate suggested reproducibility practices from the literature into quantum software experiment tracking and identify aspects of how experiment tracking improves quantum software development.
    \item We demonstrate MLflow adaptation for experiment tracking in quantum software development using a realistic, noisy, and error-mitigated quantum reservoir computing model for time-series prediction.
\end{itemize}

The rest of the article is organized as follows. In Section~\ref{background}, we introduce the background of the article. In Section~\ref{case}, we present our case study. In Section~\ref{insights}, we provide the insights we obtained during the study. Finally, in Section~\ref{conclusions}, we present our conclusions.

\section{Background on experiment tracking in quantum software development}\label{background}

In this section, we present a high-level background to various dimensions that quantum software development has and how experiment tracking has been utilized in the field to date. Many of these dimensions are later concretely tracked in the case study. As quantum computing hardware has advanced, the need for quantum-specific software engineering has emerged to recognize and address problems arising from this novel paradigm~\cite{piattini2022quantum}. 
Modern classical software development has inspired workflows for quantum software development, taking into consideration the specifics of the NISQ (Noisy Intermediate-Scale Quantum) era of quantum computing~\cite{dwivedi2024quantum}. 
One such development model, the quantum software development life cycle, specifies the internal quantum-specific processes, separating the process into the quantum workflow lifecycle, the classical software lifecycle, and the quantum circuit lifecycle~\cite{weder2022quantum}.

\subsection{Lifecycles in quantum software development}

Next, we examine how different development and programming models account for circuit readiness in the development process and how they suggest analyzing outputs and execution quality. 
In their 2022 article, Weder et al.~\cite{weder2022quantum} presented a model in which the software development lifecycle is divided into three sub-processes: the quantum workflow lifecycle, the quantum circuit lifecycle, and the classical software lifecycle. 
From this process model, we focus on quantum algorithm and circuit development, leaving us with several actions in the quantum circuit lifecycle and an analysis step in the workflow lifecycle, overlapping in critical parts with what we present as the experiment tracking and analysis process. 

Quantum circuit lifecycle steps considered as part of quantum software development in the model are: hardware-independent implementation - programming the logical circuit, testing - including verification, circuit enrichment - including state preparation and oracle expansion, hardware selection, optimization and compilation, execution, and error mitigation. 
In their 2024 article, Pérez-Castillo et al.~\cite{perez2024guidelines} presented an incremental commitment spiral model for quantum-classical systems. 
The article's main focus is on the broader perspective of information system development, rather than on the internals of quantum circuit design. 
From that perspective, it emphasizes key elements of the quantum software development phase, namely hardware-software co-design, monitoring changes, and continuous verification and validation, yet it does not go into detail on evaluating readiness or execution quality. 
Other articles addressing the subject consider both Agile~\cite{khan2024agile} and DevOps-like~\cite{gheorghe2020quantum} methods for quantum development, and share the core idea of circuit development as a single internal process, without further elaboration. While these articles offer limited insight into the details of quality evaluation, they agree on the process steps and flow.

\subsection{Diversity and quality of quantum computing ecosystems}
   
In recent years, many quantum computers have become publicly accessible to users. 
The hardware is available to users through cloud services offered by hardware companies that provide access to their devices on proprietary cloud platforms, as extensions to classical cloud services such as Amazon's AWS or Microsoft Azure, and as part of HPC systems~\cite{nguyen2024quantumcloudcomputingreview}. 
The available hardware specifications vary greatly depending on the platform and provider, and even across different devices within the same platform. 
The differences stem from the underlying qubit technologies, which vary across superconducting, trapped ion, photonic, and neutral-atom platforms, and extend through each software layer, including available SDKs and supported programming languages. 
Some of these differences are determined by the qubit implementation technology in question, while others are determined by design choices, e.g., chip architecture or native gates. 
Even within the same qubit technology, the implementations differ from each other~\cite{abughanem2025superconducting, de2021materials}. 
This article will focus only on superconducting devices and their features, as our experiments are targeted to one, yet for the most part, the principles apply to all current quantum computers.

Ideally, when a developer works at a higher abstraction level, creating a circuit using a gate-based programming model, these underlying differences should not matter.  Unfortunately, with current NISQ devices, this is not the case. Execution quality varies widely across execution platforms and even across different runs on the same device. To address internal issues that affect the transpiled code and, consequently, execution, we will explore the most important elements directly affected by the selected target hardware. 

On a smaller scale and with simulated execution, verification of the results is simpler and has multiple approaches to it, enabling the evaluation of the results accordingly. For example, small-scale simulations, up to 20-30 qubits, circuits allow debugging, following classical methods such as assertions and step-by-step execution~\cite{lewis2023formal, di2024need}. When moving to quantum hardware execution, the no-cloning theorem and state collapsing in measurement will prevent the developer from using such methods, setting demands for data to be collected and further evaluating the obtained results.

Some level of errors in quantum computing is inevitable, and different errors arise from various sources, such as decoherence within the system or the environment interacting with the computer, from the measurement process, and even from initializing qubit states before computational operations. 
Therefore, in the development towards quantum advantage, error mitigation and correction will play a key role by enabling quantum software to operate in the noise levels present in hardware, both currently and in the foreseeable future~\cite{devitt2013quantum}. 
Yet, with current NISQ hardware, we are not within the error threshold required to implement known error-correcting codes on hardware. 

For devices with current error levels, we must rely on error mitigation methods to improve the outcomes delivered by the hardware.
This is also a key motivation to include an advanced error-mitigation element in the case study.
Several effective approaches are known to mitigate errors in the current system. 
A useful error mitigation method should demand modest qubit overhead, guarantee its accuracy within named boundaries, be practical to apply in experiments, and have few or no assumptions about the computations on which they are applied. 
Yet all known error mitigation methods have scaling limitations and offer improvements only within certain bounds on the number of qubits and/or the number of gates~\cite{cai2023quantum}.
These limitations in error mitigation in the current era will require monitoring by the developer of their performance and effectiveness, and evaluation of the cost of performing them against the improvement gained.

For each quantum computer, the quality and performance of each qubit are recorded, monitored, and logged. 
As hardware performance changes over time, data should be collected and updated for each calibration of the device. 
This data should then be available to the developer as the \textit{quality metric set}. 
From a broader perspective, execution quality may be considered a subset of quantum hardware benchmarks. 
To thoroughly evaluate the hardware, a wide range of benchmarking procedures and tools is available to hardware developers and users. 
Effective benchmarking must be well-motivated to measure relevant performance metrics, well-defined to ensure unambiguity, implementation-robust to prevent exploitation, and system-robust, efficient, and technology-independent~\cite{proctor2025benchmarking}.
The exact calibration metrics set may vary slightly between hardware operators and system providers, but, based on the same metrics and similar benchmarks, the system aims to provide users with data on device performance, according to the latest calibration and benchmark results. The core metrics used for QPU characterization are T1 and T2 times, gate error rates, and readout error rates. T1 - relaxation time describes the time it takes for each qubit to lose the energy to hold the excited state $\ket{1}$ or to gain energy enough to lose the state $\ket{0}$. T2 - dephasing time describes the loss of coherence in the superposition state. In other words, as qubits and their state are often presented on the Bloch sphere, T1 presents relaxation time on the longitudinal axis and T2 on the horizontal axis~\cite{de2021materials, deng2024calibration}. 
Gate error rate or gate fidelity describes how well the physical implementation of a given gate performs the aimed transformation on the qubit, or qubits. Gate fidelities are often measured separately for single-qubit gates and two-qubit gates. Gate fidelity is measured as a percentage, allowing users to estimate the number of gates that can be applied in a circuit while still performing as intended. Finally, readout errors, or readout fidelity, are values that describe errors occurring during the measurement of each qubit~\cite{deng2024calibration, Krantz_2019}.
  
\subsection{Experiment tracking}

Log tracking in software systems provides information about system actions, enabling developers to monitor software behavior. In classical software development, multiple approaches and practices have been proposed for logging data from a system. Some focus more on finding errors or anomalies in execution, while others are focused on quality assurance, reliability, or dependability~\cite{batoun2024literature, candido2021log}. Different software domains demand different logging practices. For example, recent developments and the experimental nature of machine learning have shown the need for experiment tracking tools and practices, in which not only exceptions or anomalies are logged, but also parameters used in each run. This data will then be used to evaluate the obtained results and to support further development~\cite{zaharia2018accelerating}. Current ML experiment-tracking solutions enable tracking of important provenance and reproducibility data in experiments~\cite{schlegel2023mlflow2prov, 11250092}. Quantum software development shares many similarities with ML development, particularly in its experimental nature, and, when conducted in the academic domain, reproducibility imposes high demands on provenance logging.

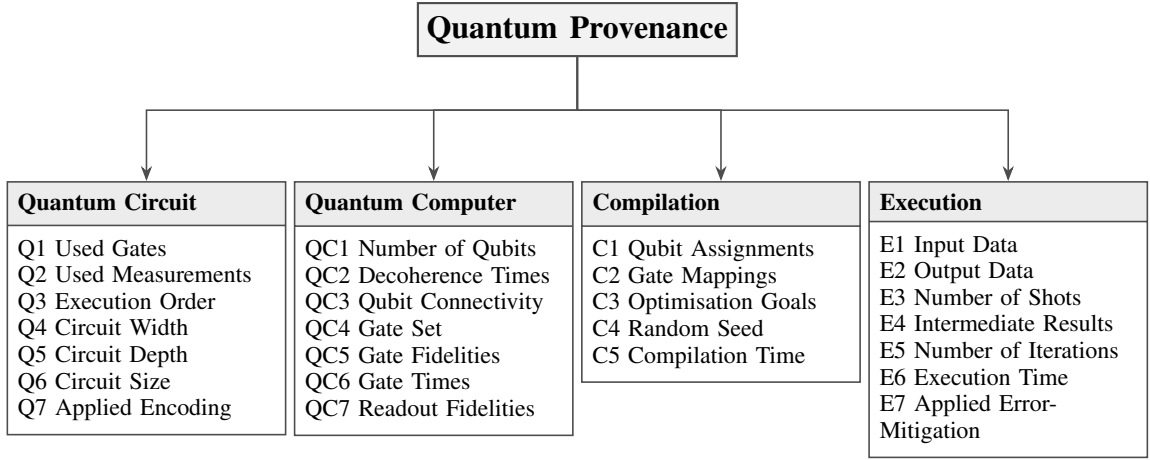
\begin{figure*}[t]
\centering
{\fontfamily{ptm}\selectfont
\begin{tikzpicture}[
    box/.style={
        rectangle split,
        rectangle split parts=2,
        draw=black!70,
        text width=3.4cm, 
        align=left,
        font=\small,
        rectangle split part fill={gray!15,white},
        rectangle split part align={center,left},
        inner sep=4pt,
        anchor=north,
        line width=0.6pt,
        minimum height=4.8cm
    },
    arrow/.style={-Stealth, semithick, black!70}
]

\node[draw=black!70, fill=gray!10, line width=0.8pt, minimum width=3.4cm, minimum height=0.7cm] 
    (root) at (0,1.5) {\large\textbf{Quantum Provenance}};

\node[box] (qc) at (-5.7,-0.5) { 
    \textbf{Quantum Circuit}
    \nodepart{two}
    Q1 Used Gates\strut \\
    Q2 Used Measurements\strut \\
    Q3 Execution Order\strut \\
    Q4 Circuit Width\strut \\
    Q5 Circuit Depth\strut \\
    Q6 Circuit Size\strut \\
    Q7 Applied Encoding\strut
};

\node[box] (qcomp) at (-1.9,-0.5) { 
    \textbf{Quantum Computer}
    \nodepart{two}
    QC1 Number of Qubits\strut \\
    QC2 Decoherence Times\strut \\
    QC3 Qubit Connectivity\strut \\
    QC4 Gate Set\strut \\
    QC5 Gate Fidelities\strut \\
    QC6 Gate Times\strut \\
    QC7 Readout Fidelities\strut
};

\node[box] (comp) at (1.9,-0.5) { 
    \textbf{Compilation}
    \nodepart{two}
    C1 Qubit Assignments\strut \\
    C2 Gate Mappings\strut \\
    C3 Optimisation Goals\strut \\
    C4 Random Seed\strut \\
    C5 Compilation Time\strut
};

\node[box] (exec) at (5.7,-0.5) { 
    \textbf{Execution}
    \nodepart{two}
    E1 Input Data\strut \\
    E2 Output Data\strut \\
    E3 Number of Shots\strut \\
    E4 Intermediate Results\strut \\
    E5 Number of Iterations\strut \\
    E6 Execution Time\strut \\
    E7 Applied Error-Mitigation\strut
};

\draw[arrow] (root.south) -- ++(0,-0.7) -| (qc.north);
\draw[arrow] (root.south) -- ++(0,-0.7) -| (qcomp.north);
\draw[arrow] (root.south) -- ++(0,-0.7) -| (comp.north);
\draw[arrow] (root.south) -- ++(0,-0.7) -| (exec.north);

\end{tikzpicture}
}
\caption{Quantum provenance taxonomy capturing key metadata across the quantum computing stack.}
\label{fig:quantum_provenance}
\end{figure*}

The majority of quantum software development is conducted for research purposes~\cite{jimenez2024quantum}, which imposes demands on the process and reporting. Conducting precise and thorough logging of the development and executions is necessary to provide detailed information on the reproducibility and integrity of the research conducted~\cite{moguel2025quantum}.
A notable effort to introduce data provenance into quantum computing is the quantum provenance framework QProv~\cite{weder2021qprov}. 
Quantum provenance categories and items are presented in Fig.~\ref{fig:quantum_provenance}. 
While the current state of hardware is largely experimental, the items mapped by QProv cover not only multiple data points on quantum execution quality, such as gate times, fidelities, and decoherence times for the given device, but also quantum circuit and compilation details. 
To collect this data, we introduced the concept of experiment tracking in quantum software development in our earlier work~\cite{kinanen2025experiment}. 
When working on quantum software and algorithm development, the developer should store the aforementioned provenance data to enable reproducibility, validation, and verification of experiments in their research, as well as to evaluate the results obtained from execution. 
Yet the experiment tracking does not limit itself to items introduced in the QProv framework or other approaches to provenance data, but should support development by enabling the evaluation of parameters and metrics in the process.

\begin{figure*}[tb]
    \centering
    \includegraphics[width=0.99\linewidth]{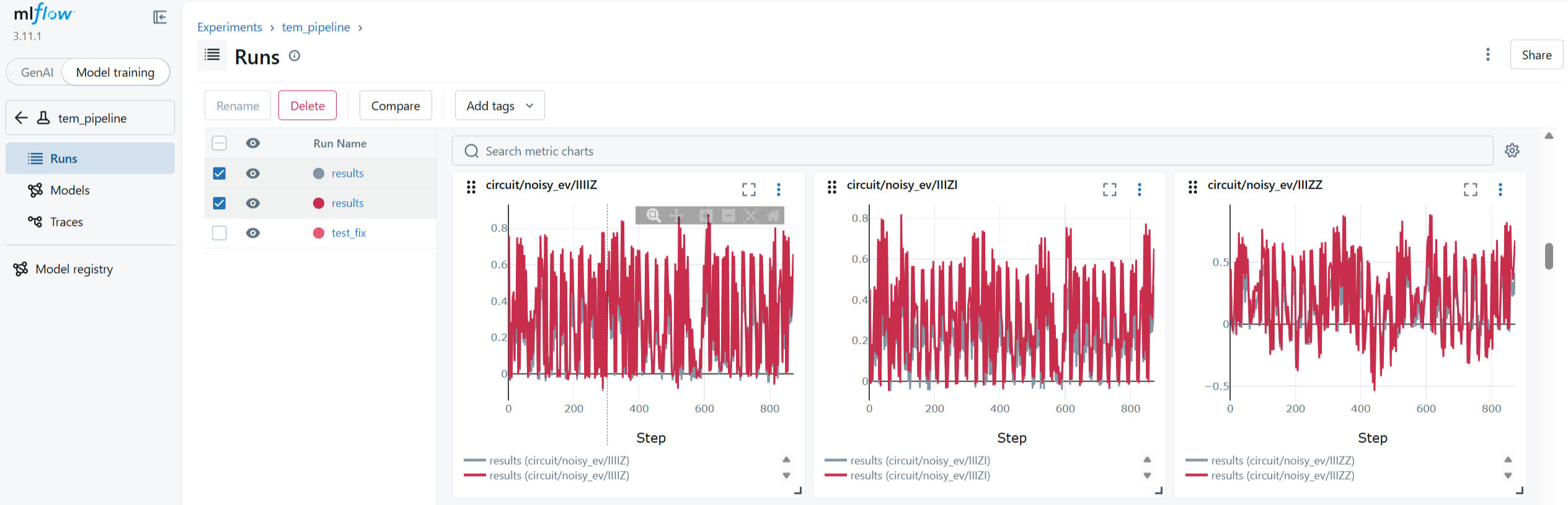}
    \caption{Example view of the graphical user interface of MLflow running in a browser}
    \label{fig:MLflow_view}
\end{figure*}

The MLflow \footnote{https://mlflow.org/docs/latest/ml/} experiment tracking tool offers several ways of managing, analyzing, and processing data obtained during the development~\cite{zaharia2018accelerating}. Although it originates from ML development, MLflow enables tracking of any necessary parameter, metric, or artifact and efficiently evaluates them, fulfilling the needs of experiment tracking in quantum software development~\cite{11250092}. An example of how MLflow enables visualization is shown in Fig.~\ref{fig:MLflow_view}.

\subsection{Related work}\label{related}

One key area of experiment tracking in scientific computations is ensuring the reproducibility of experiments. In quantum computing, several areas have distinct requirements in this context. As quantum hardware is accessed through cloud services with vendor-controlled access, there might be queues to access the same device due to other users, and the configurations are constantly changing. Given the limitations, it is necessary to provide a detailed description of the hardware used and its configuration at the time of the experiments~\cite{mauerer20221}. 

As introduced in Section~\ref{background}, Weder et al.~\cite{weder2021qprov} have presented a framework and software, called QProv, to enable users to collect provenance data in quantum computing experiments. Yet the software does not provide the extended tools to allow experiment tracking and logging outside these set provenance items, or evaluating execution results or model accuracy in machine learning use cases.

In the article on reproducibility and an exploratory study on quantum software experiments, Gierisch et al.~\cite{gierisch2025qef} introduced their approach to the problem -- Quantum Experiment Framework (QEF). QEF aims to support experimental researchers by enabling them to evaluate their work in a structured manner. A key difference between the QEF and our solution and the practice presented is that QEF uses the algorithms and evaluation metrics provided in the framework, whereas, as introduced, MLflow allows the developer to define their own metrics, parameters, and artifacts to track. For the same reason, QEF will not be suitable for a variety of quantum computing experiments outside the scope or for moving towards production-level software. 

Among the research-based approaches to the topic, the open-source software project Aqueduct\footnote{https://aqueducthub.github.io/aqueductcore/main/} has been working on a related topic. Aqueduct is a quantum experiment management platform that aims to enable a similar experiment-tracking process to what we conducted on MLflow. Yet the project has been archived with the latest release in 2024.

In our previous work~\cite{kinanen2025toolchain}, we proposed an iterative model for quantum circuit development that guides developers in using a specific type of hardware and scaling quantum execution to suit the platforms available in the development environment. 
The process evolves from initial local device execution (e.g., a laptop with a CPU) to more efficient GPUs or CPUs (e.g., in an HPC or datacenter environment), and finally, when the program has reached maturity on a smaller scale, to QPU execution. 
Following this model not only increases demand on development tooling and methods, but also requires educated decision-making between iterations, considering correctness and maturity before moving to quantum execution. 

Among provenance and experiment management for quantum software, a complementary line of research addresses the quality of quantum software through measurable properties. Recent work has proposed models for quantum software quality attributes, such as an analyzability model for quantum programs validated through a series of experiments~\cite{diazmunoz2026analyzability}. Data collected with systematic tracking practices as presented will enable following the suggested models and provide the necessary data.


\section{Case study: time series prediction with quantum reservoir computing}
\label{case}










In this section, we present a technical case study demonstrating how experiment tracking and logging can improve a realistic quantum software development pipeline, built on quantum reservoir computing and tensor-network error mitigation. Quantum machine learning (QML) is a broad term describing a variety of approaches that link to quantum computation and machine learning. The approaches may be divided into sub-areas depending on the type of data used and the type of algorithm. In QML applications, common challenges include selection and preprocessing of data sets, embedding schemes and data encoding, training landscapes, measurement schemes, and noise in current hardware~\cite{cerezo2022challenges}. The most commonly known methods in QML fall into sub-categories of variational quantum classification algorithms, quantum support vector machines, and quantum neural networks. Numerous theoretical and practical proposals are introduced throughout the field, seeking the advantages of quantum methods\cite{rodriguez2025survey}. 

\begin{figure}
    \centering
    \begin{tikzpicture}[
        node distance=0.6cm and 0.8cm,
        box/.style={rectangle, draw=black!75, rounded corners=3pt,
                    text width=7cm, minimum height=0.8cm,
                    align=center, font=\small, inner sep=5pt,
                    line width=0.6pt},
        arrow/.style={->, thick, >=stealth, draw=black!75}
    ]

    \definecolor{oi_blue}{RGB}{86, 180, 233}
    \definecolor{oi_green}{RGB}{0, 158, 115}
    \definecolor{oi_orange}{RGB}{230, 159, 0}
    \definecolor{oi_vermillion}{RGB}{213, 94, 0}

    \node[box, fill=oi_orange!25, draw=oi_orange!80] (step1)
        {1. Classical time series data \& workload preparation~\cite{valterUo_mackey_glass_qr}};

    \node[box, fill=oi_vermillion!25, draw=oi_vermillion!80, below=of step1] (step2)
        {2. Workload execution on a noisy simulator~\cite{valterUo_povm_helmi_container}};

    \node[box, fill=oi_blue!25, draw=oi_blue!80, below=of step2] (step3)
        {3. Error mitigation classically~\cite{valterUo_tem_q50_container}};

    \node[box, fill=oi_green!25, draw=oi_green!80, below=of step3] (step4)
        {4. Reservoir computing's classical readout phase~\cite{valterUo_mackey_glass_qr}};

    \draw[arrow] (step1) -- (step2);
    \draw[arrow] (step2) -- (step3);
    \draw[arrow] (step3) -- (step4);

    \end{tikzpicture}
    \caption{Summary of use case actions. References refer to the GitHub implementations.}
    \label{fig:use_case_summary}
\end{figure}
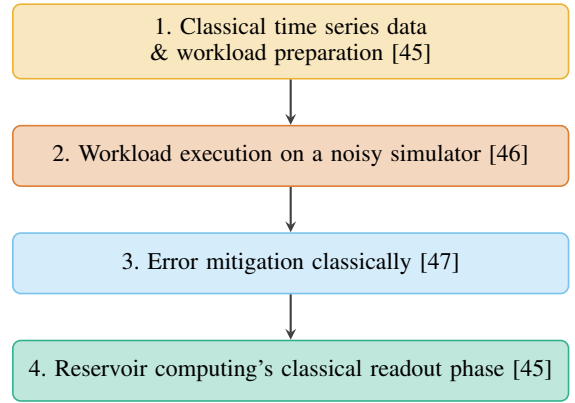

The case study pipeline comprises four phases, each of which logs multiple metrics, parameters, and artifacts that the developer may wish to track. The implemented stages in the model are depicted in Fig.~\ref{fig:use_case_summary}. The selected case study is quantum reservoir computing applied to conventional time-series prediction with chaotic data. Quantum reservoir computing is a non-variational quantum machine learning paradigm in which a quantum computer is used as a reservoir to map low-dimensional data into a higher-dimensional space~\cite{PRXQuantum.3.030325,Kornjaca_Hu_Zhao_Wurtz_Weinberg_Hamdan_Zhdanov_Cantu_Zhou_Bravo_et}. This way, one can ``linearize'' the non-linear data and use classical linear models for regression on the higher-dimensional data. The expected utility and improvement is that quantum computers, combined with classical linear models, might be a faster and better combination for non-linear regression than non-linear classical models alone.

The error mitigation we employ is the so-called tensor network error mitigation (TEM) algorithm~\cite{filippov2023scalable}. The algorithm requires performing state tomography with so-called Positive Operator-Valued Measures (POVM)~\cite{Uotila_2025} and inverting the observed noise channels. We use a noisy five-qubit simulator with relatively high depolarizing noise in single- and two-qubit gates. This TEM-based error mitigation pipeline has been shown to be optimal in a certain sense and offers several features that make it one of the most promising error mitigation algorithms~\cite{filippov2024scalability}. It is offered as part of premium IBM Quantum functions. Since this implementation is not publicly available, our custom TEM implementation is based on Filippov et al.~\cite{filippov2023scalable}. We emphasize that this implementation likely differs from any production-ready implementation of the algorithm. Nevertheless, the quantum reservoir model with TEM provides a realistic, technically advanced pipeline that serves our case study well. 

We perform each step (1.--4.) in Fig.~\ref{fig:use_case_summary} separately because this makes debugging, logging, and overall development easier. Each stage outputs logged data that serves as input to the next stage. The data can be stored and retrieved with MLflow, making logging a substantial element of the studied QML pipeline. The execution code for the noisy simulator and the tensor network error-mitigation modules operates independently of the other stages and generalizes to any workload that consists of circuits and observables, and we have also designed it to work with current real hardware. We next describe the case study in detail.

\subsection{Classical time series and workload preparation}

In this subsection, we describe stage 1, which comprises the classical time-series data preparation for the quantum reservoir computing model. At this stage, we use the Mackey-Glass time-series data, which is a widely used in classical reservoir computing~\cite{doi:10.1126/science.267326}. The time series is defined with the Mackey-Glass delayed differential equation as
\begin{equation*}
    \frac{dx(t)}{dt} = \frac{ax(t-\tau)}{1+x(t-\tau)^n} - bx(t),
\end{equation*}
where we choose the default instance with parameters $\tau = 17$, $a = 0.2$, $b = 0.1$, $ x_0 = 1.2$, and $ n = 10$. We generate $180$ data points and rely on the implementation in \texttt{reservoirpy}~\cite{ReservoirPy}.

After generating the Mackey-Glass time-series data, we min-max scale the values and then construct sliding windows of size $5$. We choose the window size to match the number of qubits in the hardware. We employ a relatively small number of qubits because the goal is to demonstrate the experiment-tracking capability, run the pipeline within a reasonable time frame, and not to focus on the scalability of the implemented method. The goal is to build a quantum reservoir computing model that learns to predict the next data point in a time series, given a window of past values. In other words, we construct a model $f$ such that
\begin{displaymath}
f(x_{i}, x_{i+1}, x_{i+2}, x_{i+3}, x_{i+4}) = x_{i+5},
\end{displaymath}
for any $1 \leq i \leq 175$.

While Mackey-Glass data are synthetic, this data preparation phase corresponds to the stage in quantum software development, or more precisely, in a quantum machine learning pipeline, where we have identified the learning task and prepared the training and test data.

The key idea in quantum reservoir learning is to use the quantum mechanical system, in this case a quantum computer following the gate-model, as a reservoir that maps the low-dimensional feature vectors into a higher-dimensional space~\cite{Mujal_2021}. In our case study, this mapping is implemented such that each data window $(x_{i}, x_{i+1}, x_{i+2}, x_{i+3}, x_{i+4})$ corresponds to a circuit. We next describe how these circuits are constructed. The key idea behind this construction is based on Quera's work~\cite{Kornjaca_Hu_Zhao_Wurtz_Weinberg_Hamdan_Zhdanov_Cantu_Zhou_Bravo_et_al_2024} but adapted from neutral atom quantum computers to the gate model. We focus on a fixed time step $t$ in the time series and consider a window size $w = 5$. Given the time series data of length $n$ as $X = (x_t)_{t=1}^{n}$, the data is divided into sliding windows as $\hat{x}_t := (x_{t}, \ldots, x_{t + w - 1}) \in \mathbb{R}^{w}$. Then, each vector $\hat{x}_t$ is mapped to a circuit generated by three Hamiltonians, as follows. The first Hamiltonian is a simple layer of Pauli-$X$ rotations:
\begin{equation*}
    H_1 = \sum_{i = 1}^{w} \frac{3}{4}\sigma_{X}^{i},
\end{equation*}
where $\sigma_{X}^{i}$ is the Pauli-$X$ acting on the qubit $i \in \left\{1,\ldots, w\right\}$. The next Hamiltonian is
\begin{equation*}
    H_2(\hat{x}_t) = \sum_{i = 1}^{w} \frac{x_i}{2}\sigma_{Z}^{i},
\end{equation*}
where $\sigma_{Z}^{i}$ is the Pauli-$Z$ acting on the qubit $i \in \left\{1,\ldots, w\right\}$. In this case, the Hamiltonian $H_2$ depends on the time series vector $\hat{x}_t$. Finally, we encode consecutive interactions in the time series as
\begin{equation}
    H_3(\hat{x}_t) = \sum_{i = 1}^{w-1} \frac{1 + 0.5x_i x_{i+1}}{4\pi} \sigma_{Z}^{i}\sigma_{Z}^{i+1},
\end{equation}
where the constants allow a favorable scaling of the coefficients. The final Hamiltonian is $H(\hat{x}_t) = H_1 + H_2(\hat{x}_t) + H_3(\hat{x}_t)$. Note that the Hamiltonian has a QAOA-circuit structure comprising a mixer layer ($H_1$), linear terms ($H_2$), and quadratic terms ($H_3$). The time evolution is then realized with unitaries of type $\exp(i\tau(H_1 + H_2(\hat{x}_t) + H_3(\hat{x}_t)))$, where $\tau$ is selected to have a discrete value ranging from $0.1$ to $2.5$ with interval $0.6$. This creates five time steps in the Hamiltonian time evolution.

Finally, the system is measured multiple times at different time steps using varying observables. At each time step $\tau$, we measure a single-qubit and two-qubit expectation values $\langle \sigma_{Z}^{i} \rangle$ and $\langle \sigma_{Z}^{i}\sigma_{Z}^{j}\rangle$, where $i$ and $j$ run over all qubits and their pairwise combinations. The measured expectation values produce a higher-dimensional embedding, which is then used in the classical training phase. In other words, the expectation values are the input for the classical linear models, such as ridge or linear regression. 

At this stage, we should have already logged many parameters to ensure the correctness of the implementation and reproducibility. The parameters that we log at this stage are listed in Table~\ref{table:params_phase_1}. We also log the following artifacts:
\begin{itemize}
    \item Circuits stored in QASM 3.0 format
    \item Observables $\sigma_{Z}^{i}$ and $\sigma_{Z}^{i}\sigma_{Z}^{j}$
    \item Virtual environment information, including Python and Qiskit versions for reproducibility
\end{itemize}

\begin{table}[tb]
\centering
\caption{Experiment parameters for preparing the time series data and circuits}
\label{table:params_phase_1}
\begin{tabular}{ll}
\hline
\textbf{Parameter} & \textbf{Value} \\
\hline
$t_{\mathrm{start}}$ & 0.1 \\
$t_{\mathrm{end}}$ & 3.1 \\
step & 0.6 \\
initial state & $H^{\otimes n}|0\rangle^{\otimes n}$ \\
readout basis & $Z$ \\
applied encoding & angle \\
number of qubits & 5 \\
dataset & Mackey-Glass \\
window size & 5 \\
seed & 42 \\
$\tau$ parameter in Mackey-Glass & 17 \\
series length & 180 \\
compilation/random seed & 42 \\
Feature vec. dim. after QRC & $75 \ (15 \text{ obs.} \times 5 \text{ points})$ \\
Total circuit count & $875$ \\
\hline
\end{tabular}
\end{table}


\subsection{Workload execution on noisy simulator}

We have developed a custom module (available on GitHub~\cite{valterUo_povm_helmi_container}) that uses the POVM toolbox~\cite{Fischer_2024} to implement classical shadows-based state tomography via positive operator-valued measurements (POVMs). Classical shadows~\cite{huang2020shadows} is a randomized measurement protocol that builds a classical description of a quantum state from a small number of randomized measurements, from which many observables can afterward be estimated classically without re-running the circuit. This property is what makes it a natural fit for this quantum reservoir computing use case, where a large number of expectation values are estimated from the same set of circuits. Our implementation allows the user to simulate a noisy quantum computer with a given noise model. We refer to Filippov et al.~\cite{filippov2023scalable} and the POVM toolbox~\cite{Fischer_2024} for details behind the state tomography method employed. The key idea is to sample a fixed number of random measurements and perform these measurements for each circuit in the workload. Since each qubit can be measured in $Z$, $X$, or $Y$ basis, we have $3^n$ options, where $n$ is the number of qubits. Each measurement is repeated a fixed number of times. For each sampled measurement, we obtain a so-called dual frame~\cite{Fischer_2024} that allows us to approximately reconstruct the quantum state from the measurement data. It is worth noting that this choice constrains the applicability of the pipeline: the number of samples required by classical shadows is governed by the shadow norm of the observables, so randomized Pauli measurements are efficient for the low-weight observables used here, whereas high-weight or global observables incur a cost that grows exponentially with their locality~\cite{huang2020shadows}. The workflow presented in this article therefore scales most effectively for models, such as this particular quantum reservoir computing implementation, whose readout consists of many local observables. In the next step, we feed this measurement data into a tensor network to mitigate noise.

At this stage, we need to keep track of information that is not tracked in classical machine learning. We again log values such as the number of qubits and the seed used. In addition to these parameters, Table~\ref{table:phase_2_params} presents some of the logged values at this stage. The most important artifacts from this stage are the file containing the results from the classical shadows method and the transpiled circuits, because these are the input for the next stage. The results contain information that allows us to approximately reconstruct the noisy state for each circuit in the quantum reservoir computing model. Some of the fields in the result data are:
\begin{itemize}
    \item Circuit identifier
    \item POVM type (Classical Shadows)
    \item POVM seed (different from the simulator's seed)
    \item PVM keys
    \item Samples
    \item Dual operators
    \item Noisy expectation values and their standard deviation
\end{itemize}

\begin{table}[tb]
\centering
\caption{Experiment parameters for performing classical shadows using POVM toolbox}
\label{table:phase_2_params}
\resizebox{\columnwidth}{!}{
\begin{tabular}{ll}
\hline
\textbf{Parameter} & \textbf{Value} \\
\hline
job file & circuits.json \\
observables file & observables.json \\
num observables & 15 \\
output file & results.json \\
measured observables & 2430 \\
shot repetitions & 2 \\
total measurements & 4860 \\
optimization level & 3 \\
max circuits per job & 10 \\
num circuits & 875 \\
backend & AerSimulator \\
noise model: type & depolarizing \\
noise model: p1 single qubit & 0.0002 \\
noise model: p2 two qubit & 0.002 \\
noise model: qubits & 5 \\
noise model: simulation method & density matrix \\
noise model: coupling map & fully connected \\
noise model: readout error & none \\
noise model: relaxation & none \\
noise model: 1q gates & rx, ry, rz, h, sx, x, y, z, s, sdg, t, tdg, id \\
noise model: 2q gates & cx, cz \\
\hline
\end{tabular}
}
\end{table}


In a realistic setting, this POVM measurement step involves an execution on real hardware. Thus, this step also includes the compilation phase. The circuits provided by the user are compiled for the selected hardware, and the resulting circuits must be stored to enable proper error mitigation in the next step. We also have to store the logical-physical qubit mapping that the router optimized, because otherwise, the error mitigation does not assign correct inverted noise channels to the executed gates. The compiled circuits are also stored as an MLflow artifact in the QASM 3.0 format. The QASM 3.0 format encodes physical qubit mapping and routing automatically, so we do not necessarily have to store the qubit mapping separately.

The current module is designed to work on the Finnish 5- and 50-qubit quantum computers. It includes a feature that allows the user to log the most recent calibration data as an artifact. In a pipeline involving real hardware, this calibration data would be used to construct the noise model, which serves as the basis for error mitigation. In practice, current quantum computing experiments are likely to span a longer time period, during which calibration occurs, resulting in varying calibration datasets. In this case, systems such as MLflow are effective because they store each experiment as a separate entity and automatically assign the appropriate timestamp and calibration data to each experiment.

\subsection{Error mitigation}

After performing state tomography, we employ tensor network error mitigation. This is a custom implementation based on the article~\cite{filippov2023scalable} and does not necessarily match the algorithm provided in the Qiskit functions~\cite{ibm_algorithmiq_tem}. Our goal is not to benchmark this algorithm but to employ it as a motivated case study, as it incorporates features that are likely to appear in future quantum computing pipelines, such as the advanced measurement scheme and novel uses of tensor networks. 

\begin{figure}
    \centering
    \resizebox{\columnwidth}{!}{
    \begin{quantikz}[column sep=0.4cm, row sep={0.7cm,between origins}]
    \lstick{$|0\rangle$} & \gate{R_z} & \gate[style={fill=orange!30}]{\mathcal{N}_1} & \gate{S} & \gate[style={fill=orange!30}]{\mathcal{N}_2} & \ctrl{1} & \gate[style={fill=orange!30},wires=2]{\mathcal{N}_3} & \meter{} \gategroup[wires=4,steps=1,style={draw=black,inner ysep=-15pt},background,label style={label position=above,yshift=0.02cm}]{POVM} & \gate[style={fill=green!30}]{D} & \gate[style={draw=purple,fill=violet!5,inner ysep=-8pt},wires=4]{\parbox{3cm}{\centering Tensor network implementing ideal gates and inverted noise channels}} & \gate[style={fill=cyan!30},wires=4]{\mathcal{O}} \\
    \lstick{$|0\rangle$} & \gate{T} & \gate[style={fill=orange!30}]{\mathcal{N}_4} & \qw & \qw & \targ{} & \qw & \meter{} & \gate[style={fill=green!30}]{D} & & \\
    \lstick{$|0\rangle$} & \qw & \qw & \gate{R_z} & \gate[style={fill=orange!30}]{\mathcal{N}_5} & \ctrl{1} & \gate[style={fill=orange!30},wires=2]{\mathcal{N}_6} & \meter{} & \gate[style={fill=green!30}]{D} & & \\
    \lstick{$|0\rangle$} & \gate{S} & \gate[style={fill=orange!30}]{\mathcal{N}_7} & \qw & \qw & \targ{} & \qw & \meter{} & \gate[style={fill=green!30}]{D} & & 
    \end{quantikz}
    }
    \caption{Overall TEM-pipeline: Every gate has its associated noise, modeled as channels $\mathcal{N}_i$. Then, we perform the POVM measurements. With the corresponding dual frame ($D$ boxes), we can map the state into a tensor network, in which the original ideal gates and inverted noise channels are applied~\cite{filippov2023scalable}. Finally, we estimate the expectation value of an observable $\mathcal{O}$.}
    \label{fig:tem_pipeline}
\end{figure}
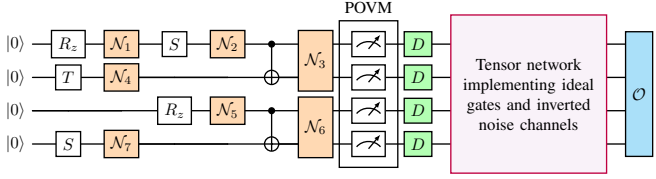

After the measurement data are logged, as described in the previous stage, we can reconstruct the quantum state for each circuit in the quantum reservoir computing model. Thus, logging automatically supports state tomography in this case study, although in the discussion section, we also identify certain data management-related issues that larger experiments might cause. The state reconstruction is performed classically by converting the dual frames into Quimb tensors, weighted by the probabilities of each measurement outcome, and representing the state as a tensor network. Based on the logged noise model at the previous step, we can construct the approximate inverse noise channels. The original transpiled gates and the corresponding inverted noise channels are appended to the tensor network, which is then contracted using Quimb. The full pipeline is summarized in Fig.~\ref{fig:tem_pipeline}. The information that enables this is all stored in the logs using experiment tracking and MLflow. For each observable, this process allows us to estimate an error-mitigated expectation value. Note that this expectation-value estimation method is favorable for the employed quantum reservoir computing model: since we estimate a relatively large number of observables, this is performed fully classically in the post-processing phase, using the same measurement data for all of them.

At this step, we again log multiple values with MLflow. Some of the logged parameters and artifacts are presented in Table~\ref{tab:phase_3}. This list is not necessarily comprehensive in the sense that tensor networks also admit many parameters, such as bond dimension, which should be tracked. Nevertheless, due to the small size of the networks in this case study, we used tensor networks without approximations.

\begin{table}[t]
\centering
\caption{TEM pipeline configuration.}
\label{tab:phase_3}
\resizebox{\columnwidth}{!}{
\begin{tabular}{ll}
\toprule
Parameter/Artifact & Value \\
\midrule
number of circuits & 875 \\
number of observables & 15 \\
POVM file & \texttt{results.json} \\
Observables file & \texttt{observables.json} \\
Calibration file & none \\
Circuits file & \texttt{transpiled\_circuits.json} \\
\bottomrule
\end{tabular}
}
\end{table}



\begin{figure*}
    \centering
    \includegraphics[width=0.95\linewidth]{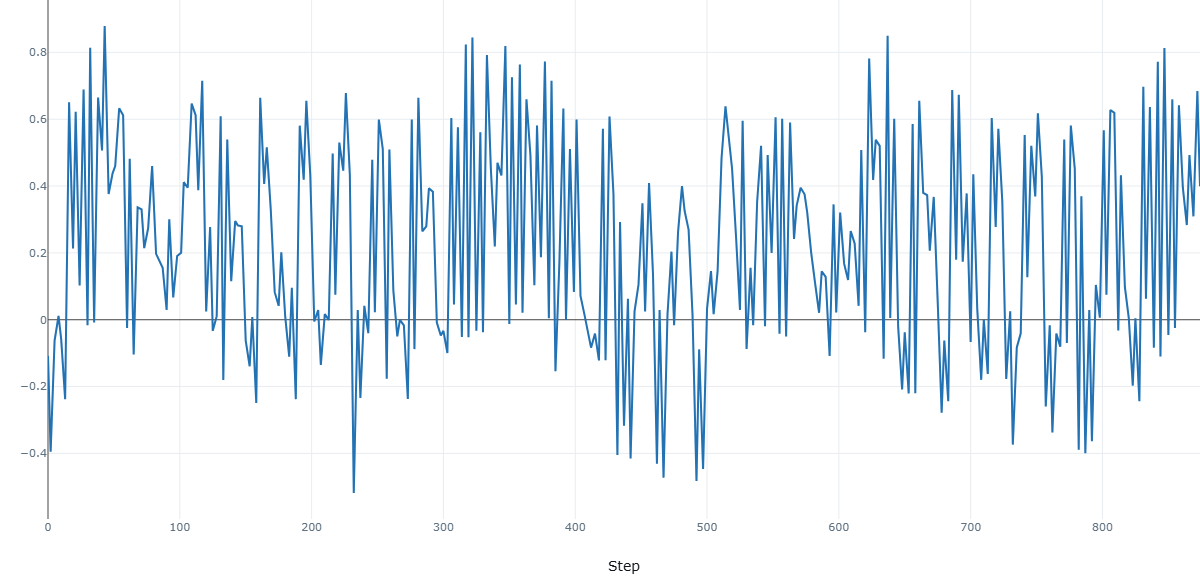}
    \caption{The error-mitigated expectation values across the 875 circuits retrieved automatically from MLflow. A step refers to a single expectation value estimation for a circuit. The y-axis contains the expectation values.}
    \label{fig:placeholder}
\end{figure*}



\subsection{Classical readout and results}

After tensor network error mitigation, we obtain error-mitigated expectation values that serve as a high-dimensional embedding of the original low-dimensional Mackey-Glass time-series data. The quantum reservoir embeddings are arguably the most important artifact from the pipeline. Thus, we present the exact code for logging them in Fig.~\ref{fig:artifact_logging}. Note that this logging mechanism saves the embeddings to the unique experiment started for this run and does not overwrite any previous data.

\begin{figure}[tb]
    \centering
    \input{figures/code_examples/artifact_logging}
    \caption{Logging quantum reservoir computed embeddings with MLflow}
    \label{fig:artifact_logging}
\end{figure}

Applying the error mitigation with tensor networks always provides estimates for the expectation values. Each expectation value estimate is accompanied by its corresponding standard deviation. We observed that this standard deviation (std) can further improve the results: values with a large std are less reliable than those with a small std. More formally, let $i = 1,\dots,N$ be the index for the samples and $j = 1,\dots,M$ be the index for observables. For each combination of sample $i$ and observable $j$, we have a TEM-mitigated expectation value $e_{ij}$ and its corresponding standard deviation $\sigma_{ij}$. For each observable, we compute the average std as $\bar{\sigma}_j = \frac{1}{N} \sum_{i=1}^{N} \sigma_{ij}$. We then perform a per-observable signal-to-noise ratio normalization (SNR) as
\begin{displaymath}
    \hat{e}_{ij} = \frac{e_{ij}}{\bar{\sigma}_j + \varepsilon},
\end{displaymath}
where $\varepsilon = 10^{-12}$. We observe that the expectation values $\hat{e}_{ij}$ further improve the results obtained from the quantum reservoir computing model. Similarly, we compute the SNR-normalized estimates of the noisy expectation values.

In this final phase, we train four linear models on four different datasets whose performance we compare: noiseless expectation values, noisy expectation values, SNR-normalized noisy expectation values, TEM-mitigated expectation values, and SNR-normalized, TEM-mitigated values. At this phase, it is relevant to log many of the values that were logged already in Table~\ref{table:params_phase_1} as well as new parameters presented in Table~\ref{tab:phase_4}. A concrete example of how simple the logging command is in MLflow is presented in Fig.~\ref{fig:param_logging}.

\begin{table}[t]
\centering
\caption{Reservoir computing's classical readout phase.}
\label{tab:phase_4}
\resizebox{\columnwidth}{!}{
\begin{tabular}{ll}
\toprule
Parameter & Value \\
\midrule
Number of cross-val. splits & 5 \\
Modes & noisy, noisy + SNR, TEM, TEM + SNR, exact \\
Classical model & Ridge \\
$\alpha$ candidates & $\{10^{-4}, 10^{-3}, 10^{-2}, 10^{-1}, 1, 10, 50, 100, 500, 1000\}$ \\
Train size & $140 \ (80\% \text{ of } 175)$ \\
Test size & $35 \ (20\% \text{ of } 175)$ \\

\bottomrule
\end{tabular}
}
\end{table}

\begin{figure}[tb]
    \centering
    \input{figures/code_examples/param_logging}
    \caption{MLflow's \texttt{log\_params} command is used to log parameters at various stages in the case study.}
    \label{fig:param_logging}
\end{figure}

Finally, we briefly present the final results obtained with the pipeline that logging supported. Performance evaluation is done using cross-validation for time series using five splits. The evaluation metric is the coefficient of determination $(R^2)$ value. An ideal model would achieve $R^2 = 1$, whereas random guessing yields $R^2 = 0$. The results, presented in Table~\ref{tab:r2_results}, show that TEM + SNR obtains results from cross-validation that are close to the exact results on average. In addition to extensive experiment logging, we consider that the results themselves are relevant. The coefficient of determination $(R^2)$ indicates that, with these parameters, the quantum reservoir computing model performs well on the task. Considering the full 80/20 training/test data, the noise does not have a big impact on the results, which is an interesting finding. This might be because the amount of training data to train the model is more than sufficient. Cross-validation results revealed that a smaller amount of training data and noisy execution decreased the quality compared to the exact simulation, which supports the idea that we had more than a sufficient amount of training samples. Interestingly, the SNR normalization technique was surprisingly effective even in the noisy simulation case.

\begin{table}[tb]
\centering
\caption{Cross-validation performance in terms of $R^2$}
\label{tab:r2_results}
\begin{tabular}{lcc}
\toprule
Mode & $R^2$ (mean $\pm$ std) & $R^2$ (full 80/20) \\
\midrule
noisy     & $-0.376645 \pm 2.3520$ & 0.9617 \\
TEM       & $-0.171476 \pm 2.0601$ & 0.9496 \\
noisy + SNR & $0.656234 \pm 0.4532$ & 0.9533 \\
TEM + SNR   & $0.735397 \pm 0.2922$ & 0.9585 \\
exact     & $0.752104 \pm 0.4378$ & 0.9980 \\
\bottomrule
\end{tabular}
\end{table}

\section{Insights obtained from experiment tracking}\label{insights}

In this section, we discuss experiment tracking and insights related to the findings in this work. We cover the impact for the end-user, data management-related issues that quantum computing pipelines create, and also point out that experiment tracking improves error tracing.






\subsection{Impact on quantum computer scientists' work}

Beyond logging, one can adopt a structured approach to quantum software development, which is currently largely research-oriented and benefits from comprehensive logs. Based on the experience in classical machine learning, experiment tracking likely produces good results also in quantum computing. Compared to classical experiment tracking, tracking current quantum environments includes elements that are not yet automatically supported, requiring manual work from users and organizations, whereas in classical ML use cases, MLflow offers automatic logging functions\footnote{https://mlflow.org/docs/latest/ml/tracking/autolog/} for all major ML Libraries. In our case study, we used the logs to adjust the amount of training data and other hyperparameters while keeping the running times reasonable. More automated logging tools would benefit users in quantum software engineering.

There exist various ways to host the tracking service. Users can either implement a tracking server themselves locally, as a cloud-hosted service within the organization, or even buy an external service. Self-hosting can be challenging and might not be suitable for individuals with no background in software engineering or computer science. Organization-level hosting removes this barrier for users and may be beneficial by allowing logs to be shared among team members. Logged artifacts might also take a massive amount of storage space, which supports organization-level hosting or using an external service hosting a tracking server. 

\subsection{Data management challenges}

The data produced by quantum computing pipelines poses new challenges for data management, which was already apparent in the deployed case study. The classical shadows pipeline already produced approximately 585 MB of measurement data, even in this small, prototypical five-qubit example. In more realistic scenarios, the amount of data will increase. Fast retrieval and integration of measurement data, as well as data such as the dual frames employed in the case study, might require new tools within the data management systems. MLflow uses a relational database; in our experiments, we used SQLite, but PostgreSQL is also supported. The relational databases are not necessarily optimized to support storing or retrieving quantum computing-based artifacts, such as measurement data and dual frames, which were stored in files. While there has been a substantial amount of research studying database optimization with quantum computing methods~\cite{Uotila_paper1,uotila2026quantum}, database-based solutions have also been used to address specific challenges in the quantum software stack~\cite{uotila2025zxdbgraphdatabasequantum,10.1145/3736393.3736694,10.1145/3722212.3725126,moflic2025ultralargescalecompilationmanipulationquantum}. This new domain will likely tackle data management challenges related to storing and retrieving data from quantum computing pipelines but the field is still taking its initial steps.

\subsection{Error tracking}

During the development of the case study, the code initially failed repeatedly, which is part of the development process. Tracking errors in code is an important aspect of experiment tracking. When the code failed in our case study, MLflow captured the error traceback information and stored it in a dedicated file, which the user can view as a result of that particular experiment.

In quantum software development, error tracking is particularly challenging because quantum computers do not support the same tracking as classical devices. This creates challenges that require new and novel methods, which support debugging quantum programs~\cite{11134329}. Previous contributions in research on debugging quantum software~\cite{di2024need,PhysRevA.89.042338} raise the question of whether MLflow can be extended to support experiment tracking within quantum programs or even quantum circuits using some form of tracking statements, whose realization could be inspired by assertions in quantum debugging research~\cite{11134329,10.1145/3307650.3322213}. Currently, MLflow is only used in classical code.

\section{Conclusions and future work}\label{conclusions}

In this article, we introduced and defined experiment tracking practices in the quantum software development context, reviewed the previous research briefly, and clarified the dimensions in which experiment tracking can be performed. These elements demonstrated that systematic experiment tracking in quantum software development admits different and often more features than classical experiment tracking. We then integrated suggested reproducibility practices from the literature into quantum software experiment tracking, which was demonstrated with a case study.

The case study demonstrated MLflow adaptation for experiment tracking in quantum software development using a realistic, noisy, and error-mitigated quantum reservoir computing model for time-series prediction. During the development, we employed experiment tracking practices borrowed from classical software and machine learning development methods. The complexity of a multi-phased development case, combined with the current state of quantum computing infrastructure, demonstrated that experiment tracking improves process quality by enabling monitoring of metrics and parameters throughout the process. MLflow also allows users to reuse collected data for experimentation and development by querying stored data. Well-conducted experiment tracking will also provide data for reproducibility in scientific experiments and research contexts. Besides tracking, the case study provided promising results.

In our case study, we have used quantum provenance data points as a starting point for mapping the necessary data to be traced. Yet the data points needed to support development often span much further and are dependent on the used algorithms, input data, and targeted hardware or simulators. In our future work, we will look deeper into mapping the tracking schema, generalizing our findings, and finding patterns to support the process. 

\bibliographystyle{IEEEtran}
\bibliography{bibliography}

\end{document}